\documentclass[aps,pra,english,showpacs,twocolumn]{revtex4}
\usepackage{latexsym}
\usepackage{bm}
\usepackage{babel}
\usepackage[T1]{fontenc}
\usepackage[latin1]{inputenc}
\usepackage{mathrsfs}
\usepackage{amsmath, amssymb}
\usepackage{graphicx}
\usepackage{ae}
\begin{document}
\title{Strong-field ionization of diatomic molecules and companion atoms:
strong-field approximation and tunneling theory including nuclear motion}
\author{Thomas Kim Kjeldsen}
\author{Lars Bojer Madsen}
\affiliation{Department of Physics and
Astronomy, University of Aarhus, 8000 {\AA}rhus C, Denmark}

\begin{abstract}
We present a detailed comparison of strong-field ionization of
diatomic molecules and their companion atoms
with nearly equal ionization potentials. We perform calculations
in the length and velocity gauge formulations of
the molecular strong-field approximation and with the
molecular tunneling theory, and in both cases we consider effects of
nuclear motion. A comparison of our results with experimental data
shows that the length gauge strong-field approximation gives 
the most reliable predictions.
\end{abstract}

\maketitle

\section{Introduction}\label{Introduction}
When a molecule is subject to an intense laser field, a series of
related strong-field processes may occur, including ionization,
dissociation and high-harmonic generation. As for atoms,
it is essential to obtain a detailed
understanding of the initial, single-ionization process in order
to describe the subsequent evolution of the system. Fully {\it ab
initio} calculations of the strong-field ionization of any molecule more
complicated than H$_2$ are impossible in the
foreseeable future. Hence, theories of general practical use
have to rely on theoretical modelling and it is the purpose of the present
work to investigate the accuracy of such models. While the
quality of {\it ab initio} calculations may be checked numerically
by studies of convergence and by identity of results in
different gauges, the quality of an approximate model has to be
checked against a more accurate model or experimental data.

In the study of strong-field ionization of molecules, three
models are widely used. These are the velocity gauge (VG)
molecular strong-field approximation (MO-SFA  VG)
\cite{muthbohm00}, the length gauge (LG) MO-SFA (MO-SFA  LG)
\cite{Kjeldsen04a}, and the molecular tunneling theory
\cite{Tong02}. The latter theory is an extension of the atomic 
Ammosov-Delone-Krainov (ADK) tunneling theory \cite{ADK}
generalized to take into account the non-spherical symmetry of
the molecular system and it is referred to as the MO-ADK theory.
The MO-SFA  VG and MO-SFA  LG theories are the velocity gauge and
length gauge versions of the Keldysh-Faisal-Reiss (KFR) atomic
theories \cite{Keldysh,Faisal,Reiss80} appropriately modified to
the molecular case.
In short, the MO-SFA theories are based on an \textit{S}-matrix formulation
where one considers the transition from a field-free initial state to a Volkov
final state, i.e., the state of a free electron in the laser field. 

To assess the quality of the models, predictions
have been compared with experimental data (see, e.g., Ref.~\cite{Tong02} and
references therein). In particular, data has been studied in detail which
compares ionization yields of diatomic molecules with yields
of atoms having nearly the same electronic binding as the molecules
-- the so-called companion atoms.
The ratio of ion signal data of diatomic molecules and their companion
atoms \cite{talebpour96, liang97, talebpour98, Guo98, DeWitt01,
Wells02} is illuminating since effects are factored out which
depend only on the binding energy. Hence a comparison of the ratio
of ionization signal of, e.g., N$_2$ and its companion Ar atom
allows one to study fairly directly effects of molecular symmetry and
ro-vibrational motion. In Ref.~\cite{Tong02} the status of
theory versus experiments was discussed: The MO-ADK theory
\cite{Tong02} gave results in satisfactory agreement with
experiments for molecules with suppressed ionization: D$_2$:Ar
\cite{talebpour98,Wells02}, O$_2$:Xe \cite{talebpour96, Guo98} and without:
N$_2$:Ar \cite{Guo98}.

For the F$_2$:Ar ratio, the MO-ADK theory predicts suppression of the
molecular signal while the experimental data \cite{DeWitt01} does
not. For the F$_2$ molecule, however, the total energy obtained
from a Hartree-Fock (HF) calculation of the molecular system is higher
than the energy of the separated atoms \cite{Helgaker00}. This
means that electron correlation effects are important for the
proper description of the binding of F$_2$, and,
accordingly, the one-electron model taken as the starting point
for the MO-ADK (and the MO-SFA) theory, is not applicable.
The models discussed above, and in the present paper, can only be
meaningfully applied on molecular systems where
effects of electron-electron correlation are small
\cite{Kjeldsen04c}.
Methods which do take electron-electron correlation into account are
presently under development \cite{Chu01,Otobe04} but these methods are not
yet applicable to molecules interacting with oscillating fields in
geometries without cylindrical symmetry.

The MO-SFA  VG  was compared with experiments in
Ref.~\cite{muthbohm00}. The theory was shown to predict suppressed
ionization of O$_2$:Xe \cite{talebpour96,Guo98} and to predict `absence' of
suppressed ionization in N$_2$:Ar \cite{Guo98}. 
In Ref.~\cite{Grasbon01}, the predictions of the MO-SFA VG
were compared with experimental above-threshold-ionization
spectra of O$_2$ ($\pi_g$ symmetry) and N$_2$ ($\sigma_g$
symmetry) and for these systems the theory gave the correct
qualitative predictions.

In general, the MO-SFA  VG approximation predicts suppressed
ionization for diatomic molecules with an antibonding highest occupied
molecular orbital (HOMO) and it
predicts no suppression of the ionization of diatoms with a
bonding  HOMO. These general predictions are based on an analysis of the
MO-SFA VG rate in terms of linear combinations of atomic orbitals where the
antibonding [bonding] rate is proportional to $\sin^2\left( \bm q \cdot
\bm R/2\right)$ $\left[ \cos^2\left( \bm q \cdot
\bm R/2\right) \right]$ and $ \bm q  \cdot \bm R \ll 1$ where $\bm q$ is the
momentum of the outgoing electron and $\bm R$ the internuclear coordinate.
For example, the MO-SFA  VG approximation
predicts suppressed ionization for O$_2$ with a HOMO of $\pi_g$ 
symmetry and no suppression of the D$_2$:Ar signal. This latter prediction
is in contrast with experimental observation \cite{Wells02} and MO-ADK theory
\cite{Tong02}, and we shall return to this discrepancy below.

In view of the above discussion at least two questions need to be
investigated in more detail. Firstly, what is the quantitative
prediction of the MO-SFA in the D$_2$:Ar case? Secondly, is it
possible to carry out a calculation in a more accurate
approximation which sheds some light on the discrepancy between
the MO-ADK calculations of Ref.~\cite{Tong02} and the
experimental results of Ref.~\cite{Wells02}? To address related questions we
have recently  developed the necessary
tools for computations in the velocity
and length gauge formulations of the MO-SFA, and in addition we
have set up a programme for the evaluation of the molecular tunneling theory
\cite{Kjeldsen04a,Kjeldsen04c}.
Our LG and VG versions of the MO-SFA are generalizations  
of the atomic adiabatic theory \cite{Gribakin97}. 
Our formulation of the MO-ADK theory 
follows Ref.~\cite{Tong02}, but our method of extracting
the angular coefficients of relevance for the evaluation of the rate
is different \cite{Kjeldsen04a,Kjeldsen04c}.
In this work, we extend our previous analysis, by taking
effects of nuclear motion explicitly into account 
and we show that the inclusion of nuclear vibrations may
lead to a significantly lower rate. 
For the molecules considered in this
paper nuclear motion is of significant importance for D$_2$ and
NO while it leads to small effects in the final results for 
the other molecules studied.

The paper is organized as follows. In Sec.~\ref{Theory}, we
describe the theories and provide a qualitative discussion of the effects of
nuclear motion. In Sec.~\ref{sec:calculation}, we give some calculational
details, in Sec.~\ref{sec:results} we present the 
results of our calculations, and in Sec.~\ref{Conclusions}, we conclude.

\section{Theory} \label{Theory}

In this section we describe the MO-SFA and MO-ADK theories and discuss how
to account for effects of nuclear motion.
In the MO-SFA one considers a state to state transition where the molecular
states are generated from moving nuclei. Contrary, in the MO-ADK there is no
detailed specification of the final state in the strong-field ionization
process,
and in order to maintain a quasistatic tunneling picture one must fix the
nuclei at an internuclear distance $R$ and 
let the electron move in the $R$-dependent potential.

\subsection{Molecular strong-field approximation including nuclear motion in
the Born-Oppenheimer approximation}
In the Coulomb gauge and in the dipole approximation the linearly polarized
laser field may be described by the vector potential
${\bm A}(t) = {\bm A}_0 \cos \omega t$,
where $\omega$ is the angular frequency. The corresponding
electric field is obtained as $\bm{F}(t) = - \partial _t \bm{A}(t)$, i.e.,
${\bm F} (t) = \bm{F}_0 \sin \omega t$.
The interaction between the field and an $N$-electron system is
[atomic units ($\hbar = m_e = e = a_0 = 1$) are used throughout]
\begin{equation} \label{eqn:Vvg}
  V^{(\text{VG})}(t) = \sum_{j=1}^N \bm{A} (t) \cdot \bm{p}_j + \frac{\bm{A}^2 (t)}{2},
\end{equation}
in the velocity gauge and 
\begin{equation} \label{eqn:Vlg}
  V^{(\text{LG})}(t) = \sum_{j=1}^N  \bm{r}_j \cdot \bm{F}(t),
\end{equation}
in the length gauge.
In either gauge, we express the
angular differential $dW/d\hat{\bm{q}}$ and total $W$ rates as
sums over $n$-photon absorptions~\cite{Gribakin97}
\begin{equation} \label{T5}
\frac{dW}{d\hat{\bm{q}}} = 2 \pi \sum_{n = n_0}^{\infty} \mid A_{\bm{q}n} \mid^2 q_n,
\end{equation}
\begin{equation} \label{T6}
W = 2 \pi \sum_{n = n_0}^{\infty} \int \mid A_{\bm{q}n} \mid^2 q_n
d\hat{\bm{q}},
\end{equation}
where the transition amplitudes corresponding to the interaction $V^{(c)}(t)$,
($c = \{\text{VG},\text{LG}\}$),
\begin{equation} \label{T7}
  A_{\bm{q}n}^{(c)} = \frac{1}{T} \int_{0}^{T} \langle \Psi_f \mid V^{(c)} (t)
\mid \Psi_i \rangle dt,
\end{equation}
involves integration over one period of the field $T = \frac{2
\pi}{\omega}$, and $\langle \mid \mid \rangle$ designates integration 
over both the electronic {\it and} nuclear
coordinates. Here  $\Psi_i$  describes the
molecular initial state and $\Psi_f$ is the final state
describing the state of the residual ion and the free electron in the laser
field.
In Eqs.~(\ref{T5})-(\ref{T6}), $n_0$ is the minimum number
of photons needed to reach the continuum, and the momentum $q_n$ is given by
energy conservation. In the Born-Oppenheimer (BO) approximation $q_n$ is
determined by Eq.~(\ref{eqn:qn}) below.

In the SFA we approximate the initial state by a
field-free molecular state.
In the BO approximation this state
is a product of an electronic state and a ro-vibrational state labelled by $\nu_i$, $J_i$. 
The electronic and vibrational states are
typically the respective ground states.
The rotational periods of the diatomic molecules are much longer than 
typical experimental pulse durations and therefore the rotational degrees of
freedom may be neglected. 
The total energy of the initial state is 
\begin{equation}
  \label{eqn:Einit}
  E_i = E_i^e(R_0) + E_{\nu_i}, 
\end{equation}
where $E_i^e(R_0)$ is the electronic eigenenergy at the internuclear
equilibrium distance $R_0$ and $E_{\nu_i}$ is the vibrational 
eigenenergy of the nuclear Hamiltonian. 
If we approximate the electronic part of the initial wave function by the
single-determinant HF wave function, the corresponding initial 
molecular wave function is
\begin{eqnarray}
  \label{eqn:initialstate}
  %  \Psi_i = \mid \psi_1(\bm r_1) \psi_2(\bm r_2) ... \psi_N(\bm r_N)
  %  \mid_{R_0} \chi_{\nu_i}(R)e^{-iE_i t}, 
  \Psi_i &=& \frac{1}{\sqrt{N!}}\det\mid \psi_1(\bm r_1) \psi_2(\bm r_2) ... \psi_N(\bm r_N)
  \mid_{R_0} \\
  & \times & \chi_{\nu_i}(R)e^{-iE_i t}, \nonumber
\end{eqnarray}
where $\chi_{\nu_i}(R)$ is the initial vibrational wave function and
the $\psi_j$'s are orthogonal single-electron wave functions.
The electronic wave function is evaluated at the nuclear equilibrium
distance $R_0$ since we assume, consistently with the BO picture,
that it will be a slowly varying function of the
internuclear distance. We have checked that the results are insensitive to
this approximation.

We seek the transition amplitude to a single-electron Volkov state
and a definite vibrational and electronic eigenstate of the molecular ion.
The application of a Volkov wave in the final state
means that the electron-ion interaction is 
neglected. Additionally, we assume that the
electronic state of the ion is unrelaxed, i.e., only the HOMO is
affected.
The final state is then
\begin{eqnarray}
  \Psi_f &=& \frac{1}{\sqrt{N!}}\det\mid \psi_1(\bm r_1) \psi_2(\bm r_2) ... \psi_V(\bm r_N,t)
  \mid_{R_0} \\
  &\times& \chi^+_{\nu_f}(R)e^{-iE_f^+ t}, \nonumber
\end{eqnarray}
where $\psi_V$ is a $(2\pi)^{-3/2}$ normalized Volkov wave function and
where the superscripts ``+''
denote the ionic state. The time-averaged energy of the
electron in the laser field is $q_n^2/2+U_p$, and the total final-state
energy is
\begin{equation}
  \label{Efinal}
  E_f = E_f^{e,+}(R_0) + E_{\nu_f}^{+} + \frac{q_n^2}{2} + U_p, 
\end{equation}
where $U_p = F_0^2/4\omega^2$ is the ponderomotive potential, and where the final state
momentum $q_n$ is determined by energy conservation $n\omega = E_f - E_i$,
i.e.,
\begin{equation} \label{eqn:qn}
  q_n = \sqrt{2 (n \omega - (E_f^{e,+}(R_0) + E_{\nu_f}^+ - E_i^{e}(R_0) -
  E_{\nu_i}) - U_p)}.
\end{equation}
The transition amplitude of Eq.~(\ref{T7}) can now be written as
($c=\{\text{VG},\text{LG}\}$)
\begin{widetext}
\begin{equation} \label{eqn:Avib}
A_{\bm{q}n}^{(c)} = \frac{1}{T} \int_{0}^{T} 
\langle \chi_{\nu_f}^{+}(R)\psi_V(\bm r_N, t)
\mid V_{N}^{(c)} (t) \mid
\psi_N(\bm r_N; R_0)\chi_{\nu_i}(R)
\rangle 
\exp{\left[i\left( E_f^{e,+}(R_0) + E_{\nu_f}^{+} - E_i^e(R_0) -
E_{\nu_i}
\right)t \right]} dt,
\end{equation}
\end{widetext}
where the $N$-electron matrix element of the one-electron operators,
Eqs.~(\ref{eqn:Vvg}) and (\ref{eqn:Vlg}), was been simplified by the
Slater-Condon rules \cite{condon30} and where $V_{N}^{(c)} (t)$ is the 
transition operator for electron $N$.
The integration over nuclear coordinates can be performed immediately to give
\begin{eqnarray}
  \label{eqn:Avib-simple}
  \nonumber
  A_{\bm{q}n}^{(c)} &=& S_{\nu_f,\nu_i}\frac{1}{T} \int_{0}^{T} 
  \int \psi^*_V(\bm r_N, t)V_N^{(c)}(t)\psi_N(\bm r_N; R_0)d\bm{r}_N
  \\
  &\times&e^{i\left( E_f^{e,+}(R_0) + E_{\nu_f}^{N,+} - E_i^e(R_0) -
E_{\nu_i}^N
\right)t} dt,
\end{eqnarray}
where $S_{\nu_f,\nu_i}$ is the Franck-Condon (FC) factor
\begin{equation}
  \label{eqn:franckcondon}
  S_{\nu_f,\nu_i} = \int \left[\chi_{\nu_f}^{+}(R)\right]^* \chi_{\nu_i}(R) dR.
\end{equation}
From Eq.~(\ref{eqn:Avib-simple}), we find the following explicit expressions for the
amplitudes
\begin{widetext}
\begin{equation}
  \label{eqn:Avib-lg} 
  A_{\bm{q}n}^{(\text{LG})} = S_{\nu_f,\nu_i}\frac{1}{T} \int_{0}^{T} \left(-E_b -
  \frac{\left[\bm{q}_n + \bm{A}(t)\right]^2}{2} \right) \tilde{\psi}_N (\bm{q}_n +
  \bm{A}(t)) \exp i \left(n \omega t + \bm{q}_n \cdot
  \bm{\alpha}_0 \sin ( \omega t) + \frac{U_p}{2 \omega} \sin (2
  \omega t) \right)t 
\end{equation}
%\end{widetext}
\begin{equation} 
  \label{eqn:Avib-vg}
  A_{\bm{q}n}^{(\text{VG})} = S_{\nu_f,\nu_i}\left(-E_b -
  \frac{q_{n}^{2}}{2}\right)
  \tilde{\psi}_N (\bm{q}_n)\frac{1}{T} 
  \int_0^T \exp i \left(n \omega t + \bm{q}_n \cdot \bm{\alpha}_0 \sin ( \omega t)
  + \frac{U_p}{2 \omega} \sin (2 \omega t)\right)t,
\end{equation}
\end{widetext}
%\begin{eqnarray}
%  \nonumber 
%A_{\bm{q}n}^{(\text{LG})} &=& S_{\nu_f,\nu_i}\frac{1}{T} \int_{0}^{T} (-E_b -
%\frac{(\bm{q}_n + \bm{A}(t))^2}{2}) \tilde{\psi}_N (\bm{q}_n +
%\bm{A}(t))\\\label{eqn:Avib-lg} &\times& \exp i (n \omega t + \bm{q}_n \cdot
%\bm{\alpha}_0 \sin ( \omega t) + \frac{U_p}{2 \omega} \sin (2
%\omega t))dt 
%\end{eqnarray}
%\begin{eqnarray} \label{eqn:Avib-vg}
%  A_{\bm{q}n}^{(\text{VG})} &=&  S_{\nu_f,\nu_i}(-E_b - \frac{q_{n}^{2}}{2})
%  \tilde{\psi}_N (\bm{q}_n)\frac{1}{T} \\ \nonumber  &\times&
%  \int_0^T \exp i (n \omega t + \bm{q}_n \cdot \bm{\alpha}_0 \sin ( \omega t)
%  + \frac{U_p}{2 \omega} \sin (2 \omega t))dt.
%\end{eqnarray}
In Eqs.~(\ref{eqn:Avib-lg})-(\ref{eqn:Avib-vg}), $\bm{\alpha}_0 = \bm{A}_0/\omega$
is the quiver radius, 
$\tilde \psi_N (\bm{q}) = (2 \pi)^{-3/2} \int d\bm{r} e^{-i\bm{q}
\cdot \bm{r}} \psi_N (\bm{r})$ is the momentum space wave function of the HOMO,
and $E_b$ is the energy difference between the final and initial state
\begin{equation}
  E_b = E_f^{e,+}(R_0) + E_{\nu_f}^{+} - E_i^e(R_0) - E_{\nu_i}.
\end{equation}
When we compare Eqs.~(\ref{eqn:Avib-lg}) and (\ref{eqn:Avib-vg}), we see that the
length gauge formulation accounts for a superimposed quiver
motion in momentum space of the bound state electron via the
presence of the $\bm{A}(t)$ term in the argument of the momentum
space wave function. Such an effect is not present in the velocity
gauge amplitude. 

To account for the Coulomb interaction between the outgoing
electron and the residual molecular ion one has typically introduced a
Coulomb correction factor. In the velocity gauge, this factor
is \mbox{$C_{\text{Coul}}^{(\text{VG})} = \left( \kappa^3/F_0\right)^{2Z_\text{ion}/\kappa}$}
with $\kappa=\sqrt{2 E_b}$ \cite{muthbohm00} and $Z_\text{ion}$ the charge of the
residual ion, while in the length gauge
\mbox{$C_{\text{Coul}}^{(\text{LG})} = \left( 2\kappa^3/F_0\right)^{2Z_\text{ion}/\kappa}$}
\cite{Perelomov66}. Both correction
factors were derived for the case of strong-field ionization of
{\it atoms}, and hence do not take into account the molecular
symmetry. In our evaluation of rates, we have found that much more
precise results are obtained in the length gauge with
$C_{\text{Coul}}^{(\text{LG})} =1$ \cite{Kjeldsen04a,Kjeldsen04c}. 
We explain the absense of a Coulomb interaction in the length gauge 
by the fact that the transition to the continuum occurs at large distances.
In this spatial region the laser-electron interaction is stronger than
the electron-ion interaction and the Volkov state is a good
approximation for the final state.
In addition to the Coulomb correction
factor, we multiply the rates by the number of equivalent
electrons in the HOMO.
Finally, to obtain the total ionization rate of the molecules, 
we must sum the contributions from
each vibrational level in the final state.
In the velocity gauge our result is equivalent to that of
Ref.~\cite{Becker01}.
For the noble gas atoms
with filled $p$ shells we sum the rates from each magnetic sub-state to
obtain the total rate of ionization.

\subsubsection{Qualitative discussion of the effect of nuclear motion}
\label{sec:vibration}

The transition amplitude of Eq.~(\ref{eqn:Avib-simple}) consists of two
factors, namely the Franck-Condon (FC) factor and an electronic matrix element and
both factors depend on the vibrational levels considered.
The rates to each vibrational level are therefore not just proportional to the FC
factors.
Instead the relative populations in the lower final vibrational states are
enhanced compared with the distribution obtained from the FC factors alone
\cite{Becker01} because the electronic matrix element is favoured by the smallest
energy differences. 
When including vibrations, the total rate summed over all final vibrational states
will therefore typically be smaller than if the vibrational ground state of the ion
had been given the weight of unity.
This latter method is nearly equivalent to fixing
the nuclei [compare Eqs.~(\ref{eqn:Avib-lg})-(\ref{eqn:Avib-vg}) with
Eqs.~(5)-(6) of Ref.~\cite{Kjeldsen04a}].

The importance of the inclusion of nuclear vibrations will depend on the
properties of the neutral molecule and the molecular ion. If their
potential curves are only shifted with respect to each other but otherwise 
exactly identical then the vibrational eigenstates will be identical too. 
The orthogonality of the nuclear wavefunctions then assures that only a
single FC will be nonzero.
We may estimate the importance of nuclear 
vibrations using molecular orbital theory and by considering the type of valence orbitals.
If the valence orbital is nearly non-bonding as, e.g., in N$_2$,
the bonding properties of the molecular ion will be approximately equal 
to the bonding properties of neutral molecule and transitions between the
vibrational ground states of the molecule and the ion dominate.
In the case of a bonding HOMO, e.g., as in D$_2$, the bonding of the 
ion will be weakened and transitions to many vibrational states will occur. 

\subsection{Molecular ADK theory}
\label{sec:adktheory}

The tunneling theory \cite{ADK,Tong02} relies on the assumption that at
any given instant of
time the system will respond to the external laser field as if it were a
static electric field.
The rate of ionization in the oscillating field may then be determined
by time-averaged static rates. Whether this
approach is reasonable or not depends on the value of the Keldysh parameter
$\gamma = \sqrt{2E_b}\frac{\omega}{F_0}$ \cite{Keldysh} with $\gamma \ll 1$
in the tunneling regime. We will only show results from the MO-ADK theory in
the intensity regions corresponding to $\gamma \leq 1$.

The tunneling rate of diatomic molecules can be determined once the
field-free asymptotic wave function is known. In a body-fixed frame,
labeled by superscript $B$ and a $z$ axis directed along the internuclear
axis, this function has the asymptotic Coulomb form
\begin{equation}
  \label{eqn:bodywf}
  \psi^B_N(\bm{r}) \sim r^{Z_\text{ion}/\kappa -1} e^{-\kappa r} \sum_{l}
C_{lm} Y_{lm}(\hat{\bm{r}}),
\end{equation}
where $m$ is the projection of the angular momentum on the internuclear
axis, and where $C_{lm}$ are expansion coefficients. 

From the asymptotic form of Eq.~(\ref{eqn:bodywf}), the total ionization rate in a static (DC)
field in the positive laboratory-fixed $Z$ direction is calculated as in the atomic
case \cite{Smirnov66,Perelomov66,Bisgaard04}, and the result is
\cite{Tong02}
\begin{eqnarray}
  \label{eqn:statrate}
  W_\text{stat}(F_0) &=&
  \sum_{m'} \frac{|B(m')|^2}{2^{|m'|} |m'|!
\kappa^{2Z_\text{ion}/\kappa-1}}\\ \nonumber &\times& \left( \frac{2\kappa
^3}{F_0}\right)^{2Z_\text{ion}/\kappa - |m'| - 1} \exp \left( -\frac{2}{3}
\frac{\kappa^3}{F_0}\right),
\end{eqnarray}
where
\begin{eqnarray}
  \label{eqn:bm}
\nonumber 
  B(m') &=& \sum_{l} (-1)^{(| m' | + m')/2} \sqrt{\frac{(2l+1)(l + | m'|)!}
{2(l-|m'|)!}}\\
&\times& C_{lm} d_{m' m}^{(l)} (\theta).
\end{eqnarray}
Here $d_{m' m}^{(l)} (\theta)$ is the middle term of Wigner's rotation
function \cite{zare} and $\theta$ is the angle between the field direction
and the internuclear axis.

In a slowly varying field, the ionization rate is found by averaging the DC
rate over an optical cycle~\cite{Kjeldsen04c}
\begin{equation}
  \label{eqn:tunnelrate2}
  W \approx \sqrt{\frac{3F_0}{\pi \kappa^3}}\frac{W^+_\text{stat}(F_0) +
  W^-_\text{stat}(F_0)}{2},
\end{equation}
where $W^\pm_\text{stat}(F_0)$ are the DC rates for the positive and negative
field directions with respect to the $Z$ direction. When the field points in
the negative $Z$ direction, Eqs.~(\ref{eqn:statrate}) and (\ref{eqn:bm})
must be modified by the substitution $C_{lm}\rightarrow (-1)^l
C_{lm}$ \cite{Kjeldsen04c}.
We see from Eqs.~(\ref{eqn:statrate})-(\ref{eqn:tunnelrate2}) that one only 
needs to know the $C_{lm}$ coefficients in order to be able to
evaluate the tunneling formula analytically. See also
Ref.~\cite{Kjeldsen04c} for a generalization of the tunneling theory to
molecules with more than two nuclei.

The tunneling theory can be extended to include effects of vibrations. In
the quasistatic picture all potentials seen by the active electron should be
regarded as being static. This means that the nuclei are fixed and the
height of the tunneling barrier will depend on the internuclear distance
chosen. We then calculate the rate for each value of $R$ and weight these
$R$-dependent rates by the $R$ probability distribution obtained from the
nuclear wave function \cite{Tong02}. As an approximate binding energy we
take the energy difference between the potential curves of the ion and
neutral molecule as calculated by HF theory. The vibrational wave
functions are constructed from the harmonic approximation to experimental 
potential curves~\cite{webbook}. 
As demonstrated in Sec.~\ref{sec:results} the effect of this $R$-dependent
weighting is quite small. Note in passing that recently vibrational
distributions were measured and calculated with tunneling
theory~\cite{urbain04}.

\section{Calculational details}\label{sec:calculation}

In order to determine the angular coefficients $C_{lm}$ we evaluate the ground
state wave function in the HF approximation. We seek an
accurate description of the orbitals at large distances with the correct
exponential behaviour. For this purpose we find in general the usual
expansion in an atomic Gaussian basis to be inadequate. Instead we solve the
HF equations fully numerically 
for the diatomic molecules \cite{kobus96} and for the atoms \cite{hf86,hf96}.
After having obtained the ground state orbitals we project the highest
occupied orbital on the spherical harmonics and match the resulting radial
functions to the form $C_{lm}r^{Z_\text{ion}/\kappa -1} e^{-\kappa r}$
treating the angular coefficients as fitting parameters. We give the $C_{lm}$'s
obtained by this procedure in Table~\ref{tab:clm}.
Since orbitals from
the HF calculation are optimal within the independent
particle approximation, the HF $C_{lm}$ coefficients should be
more accurate than the multiple scattering coefficients reported in 
Ref.~\cite{Tong02}.

The knowledge of the coefficents $C_{lm}$ is sufficient to evaluate
the MO-SFA LG rate accurately \cite{Kjeldsen04c}. For the evaluation of the 
MO-SFA VG transition amplitudes, we make a numeric Fourier transform of the HOMO.

If we use the experimental ionization potential in the calculation of
$\kappa$, the HOMO is not guaranteed to follow the correct asymptotic
form of Eq.~(\ref{eqn:bodywf}). The asymptotic form will be similar
but with $\kappa^\text{HF} = \sqrt{2 |\varepsilon_\text{HOMO}|}$
substituted for $\kappa$ with $\varepsilon_\text{HOMO}$ the eigenvalue
of the one-electron HF (Roothaan) equation. For the highly correlated F$^-$ ion,
for example, 
the two different values of $\kappa$ differ by 20\% and the
application of a wrong asymptotic wave function will introduce a large error
when calculating the MO-SFA LG ionization rates \cite{gribakin00}. 
For the systems considered here the largest difference between 
the experimental $\kappa$ and $\kappa^\text{HF}$ is 10\% for NO 
and less than 5\% for the remaining systems. Thus we conclude that
correlation effects do not affect the outer electron significantly for the
systems considered here. We believe that the smallness of the differences
between $\kappa$ and $\kappa^\text{HF}$ justifies 
the application of the single-active electron models.

\begin{table}
    \centering
    \caption{The molecular and atomic properties necessary for the
    evaluation of the present MO-SFA LG and the MO-ADK theory. 
    $I_p$ is the experimental
    adiabatic ionization potential and $R_0$ is the equilibrium
    distance~\cite{webbook}.
    Furthermore we give the
    angular coefficients $C_{lm}$ from our Hartree-Fock based calculation 
    together with the coefficients from Ref.~\cite{Tong02}. We
    have chosen the origin at the geometrical midpoint. Franck-Condon factors 
    and vibrational energies can be found in the references indicated after
    each molecular species.}
    \begin{ruledtabular}
      \begin{tabular}{llrrrrrrrr}
&        &$I_p$     &$R_0$ &$C_{0m}$&$C_{1m}$&$C_{2m}$&$C_{3m}$&$C_{4m}$& \\
&        &      (eV)&      (\AA)&        &        &        &        &        &    \\
     \hline 
D$_2$ ($\sigma_g$)&\cite{dunn66}&15.47&0.742& 2.44 &    &0.14  &    &0.00  &\\
&&     &     &  2.51&    &0.06&    &0.00& \cite{Tong02}\\
N$_2$ ($\sigma_g$)&\cite{halmann65}&15.58&1.098&3.46  &    &1.64  &    &0.12  &\\
&&     &     &2.02&    &0.78&    &0.04&\cite{Tong02} \\
O$_2$ ($\pi_g$)  &\cite{nicholls68} &12.03&1.208&      &    &1.04  &    &0.07  &\\
&&     &     &      &    &0.62&    &0.03&\cite{Tong02} \\
S$_2$ ($\pi_g$)& \cite{Berkowitz}\footnote{Based on photoelectron spectrum}
& 9.36&1.889&      &    &1.46  &    &0.24  &\\
&&     &     &      &    &0.81&    &0.07&\cite{Tong02}\\
\hline
CO ($\sigma$) &\cite{nicholls68}&14.01 &1.128&-3.93 & 2.79 &-1.59  &0.31  &-0.09&\\
&&      &     &1.43&0.76&0.28 &0.02&     &\cite{Tong02}\\
NO ($\pi$)  & \cite{nicholls68} & 9.26 &1.151&      &-0.25 & 0.82  &-0.06 & 0.04&\\
&&      &     &      &0.22&0.41 &0.01&     &\cite{Tong02}\\
SO ($\pi$)  &\cite{dyke97}\footnotemark[1]  &10.29 &1.481&      & 1.09 &-1.25  & 0.34
&-0.12&\\
&&      &     &      &0.41&-0.31&0.01&     &\cite{Tong02}\\
\hline
Ar ($p$)&      & 15.76&     &      & 2.51 &    &    &     &\\
&&      &     &      &2.44&    &    &     &\cite{Tong02}\\
Kr ($p$)&      & 14.00&     &      & 2.59 &    &    &     &\\
&&      &     &      &2.49&    &    &     &\cite{Tong02}\\
Xe ($p$)  &    & 12.13&     &      & 2.72 &    &    &     &\\
&&      &     &      &2.57&    &    &     &\cite{Tong02}\\
    \end{tabular}
    \end{ruledtabular}
    \label{tab:clm}
\end{table}

With the ionization rates at hand, we calculate the ion signals by
integrating over the temporal and spatial profile of the pulse
\cite{Kjeldsen04c,fittinghoff93}. 
In all experiments discussed here, the ions were collected through a
small pinhole near the beam waist and the spatial integration should be
restricted to this region. 
As the signals saturate, the signal ratios will approach unity for such a setup. 
We calculate the rates and signals for each molecular
orientation and average the signals over 
orientations in order to simulate randomly oriented ensembles.

\section{Results and discussion}\label{sec:results}

\subsection{N$_2$:Ar}\label{sec:n2ar}

Several experiments have been performed on N$_2$ and its
companion atom Ar \cite{liang97,Guo98,Grasbon01,DeWitt01}, and there are significant
disagreements between the results. The yield
ratios were estimated to be N$_2$:Ar $\approx$ 0.7 \cite{Guo98}, 0.2
\cite{liang97}, 1 \cite{Grasbon01} and 1.7 \cite{DeWitt01}, respectively. 
Previous theoretical calculations with the MO-SFA VG model predicted the
N$_2$:Ar ratio to be above unity \cite{muthbohm00} while MO-ADK
calculations predicted the ratio to be $0.4-0.6$.
In Ref.~\cite{Tong02} it was shown that the experimental differences could
not be explained by differences in pulse lengths and the possibility of dynamical
alignment. 

\begin{figure}
  \begin{center}
    \includegraphics[width=0.9\columnwidth]{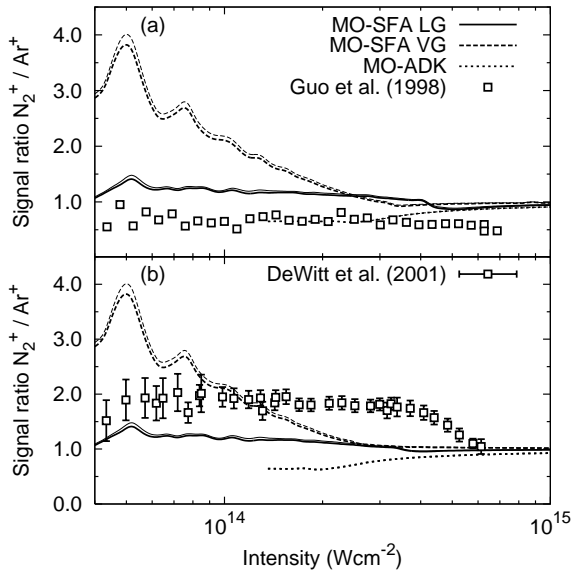}
  \end{center}
  \caption{Intensity dependent ratios between the yields of N$_2$ and Ar
  ions. In both panels the laser wavelength is $800\, \mbox{nm}$. In panel (a), the pulse
  duration is $30\, \mbox{fs}$ and the experimental data is from Ref.~\cite{Guo98}.
  In panel (b), the pulse duration is $100\, \mbox{fs}$ and
  the experimental data is from Ref.~\cite{DeWitt01}. 
  The thin lines are calculations with fixed nuclei and the thick 
  lines are the corresponding calculations including vibrations.
  }
  \label{fig:n2ar-ratios}
\end{figure}
In Fig.~\ref{fig:n2ar-ratios} we show our calculations for the N$_2$:Ar
ratio. First, we note that the results from each model are nearly
independent of pulse duration.
In the length gauge we predict the ion yields of N$_2$ and Ar to be 
nearly identical. These findings are in reasonable agreement with the experimental
data from Refs.~\cite{Guo98,DeWitt01}.
The differences between our MO-ADK results and the
results of Ref.~\cite{Tong02} lie in the values of the angular $C_{lm}$
coefficents. We find the ratio to be slightly lower than unity
in agreement with Ref.~\cite{Guo98} [Fig.~\ref{fig:n2ar-ratios}~(a)].
The velocity gauge MO-SFA predicts the ion yield of N$_2$ to be somewhat
larger than for Ar in accordance with Ref.~\cite{muthbohm00}.
We find an oscillating behaviour of the signal ratio in the velocity
gauge which is not supported by the experimental data.
\begin{figure}
    \includegraphics[width=\columnwidth]{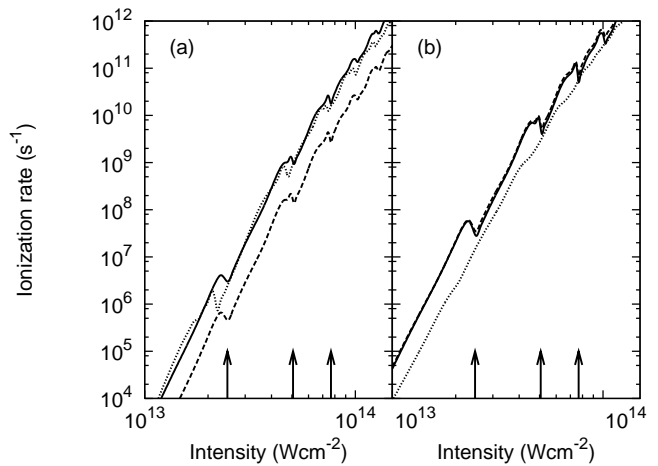}
  \caption{The ionization rates of N$_2$ and Ar (dotted) at intensities around the
  first channel closings obtained from (a) MO-SFA LG and (b) MO-SFA VG.
  For N$_2$, we give the rates at parallel 
  (solid) and at perpendicular geometry (long-dashed). 
  The wavelength is $800\, \mbox{nm}$ and the intensities at which the
  channels close for N$_2$ are indicated by the arrows. Ar has a slightly
  higher ionization potential and the corresponding intensities lie at bit
  lower. }
  \label{fig:channelclose}
\end{figure}
The origin of this artifact lies in the concept of channel closing:
At low intensities continuum-continuum transitions are very weak. The rate
of ionization is then typically dominated by the rate which originates from the lowest
number of photon absorptions, $n_0$. As the intensity increases the ponderomotive
potential rises and eventually leads to the closing of the $n_0$
channel. The effect of channel closing on the total ionization rate depends
on the relative importance of the $n_0$ process compared with the excess-photon 
processes. As the intensity increases and a channel closing approaches, 
the momentum corresponding to $n_0$ photon absorptions will become small.
If the HOMO does not contain an $l = 0$ component, then $\tilde{\psi}_N \rightarrow 0$
as $q \rightarrow 0$, and for such systems the MO-SFA VG amplitude,
Eq.~(\ref{eqn:Avib-vg}), will be suppressed. Near the
channel closings the contribution from the $n_0$ channel is thus small and
the total rate will not be significantly affected when this channel closes.
If on the other hand the HOMO contains an $l = 0$ component, the low momenta will
generally be favoured and the $n_0$ channel gives a large contribution of the
total rate. In this case, the closing of a channel will therefore lead to an
abrupt decrease of the total rate.
These differences were previously mentioned in Ref.~\cite{jaron04}.
In the length gauge, the $n_0$ channel will be important regardless of the
type of orbital due to the precense of $\bm A(t)$ in the argument of the
momentum space wave function [see Eq.~(\ref{eqn:Avib-vg})] and we always 
find abrupt decreases of the rate across a
channel closing.
We show the various effects described above in Fig.~\ref{fig:channelclose}.
The HOMO of N$_2$ is a $\sigma_g$ ($l = 0$ component)
orbital and correspondingly we find a local minimum in the rate around the channel
closings. The highest occupied atomic orbitals of Ar are the degenerate
$p$ ($l=1$) orbitals, and we find local minima in the length gauge rate while the
velocity gauge rate increases smoothly near the channel closings. The
differences between the parallel and perpendicalar geometries of N$_2$ in the two
gauges was explained in Ref.~\cite{Kjeldsen04a}.

\subsection{D$_2$:Ar}\label{sec:d2ar}

Two experiments have been reported on D$_2$ and Ar
\cite{talebpour98,Wells02} and both showed a suppression of the
D$_2$ signal compared with the Ar signal.
As mentioned in the introduction the MO-ADK calculations
reproduced this result \cite{Tong02} while 
the MO-SFA  VG model predicts absence of suppression because the HOMO of
D$_2$ is a bonding $\sigma_g$ orbital.

\begin{figure}
  \begin{center}
    \includegraphics[width=0.9\columnwidth]{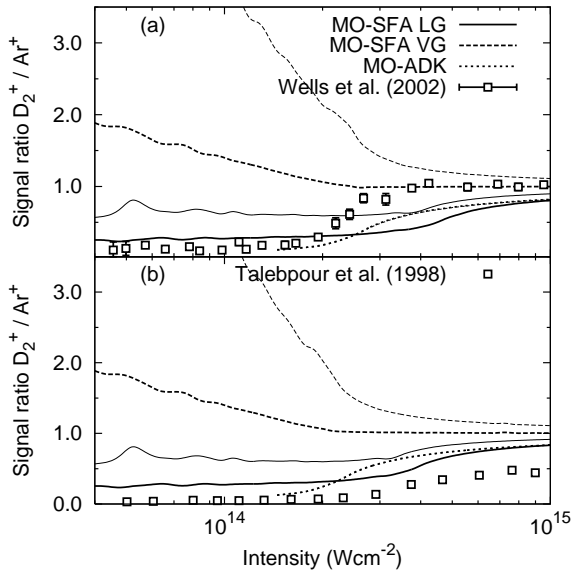}
  \end{center}
  \caption{Intensity dependent ratios between the yields of D$_2$ and Ar
  ions. In both panels the laser wavelength is $800\, \mbox{nm}$. 
  The thin lines are calculations with fixed nuclei and the thick 
  lines are the corresponding calculations including vibrations.
  In panel (a), the pulse
  duration is $100\, \mbox{fs}$ and the experimental data is from 
  Ref.~\cite{Wells02}. In panel (b), the pulse duration is $200\, \mbox{fs}$ and
  the experimental data is from Ref.~\cite{talebpour98}.
  }
  \label{fig:d2-ar-ratio}
\end{figure}
In Fig.~\ref{fig:d2-ar-ratio}, we show our calculations and the experimental
sets of data. We do indeed find absence of suppressed ionization when we
use the MO-SFA VG model. On the other hand the MO-SFA LG and MO-ADK
models both correctly predict suppression. 
At high intensities the ratio will approach unity as both
signals saturate. The MO-SFA LG model has some problems in predicting the
correct intensity at which this saturation occurs, but we note that 
the experiments do not
agree on the saturation intensity either. The longer pulse length should
be equivalent to a lower saturation intensity but this is clearly not the
case when comparing the experimental data in Figs.~\ref{fig:d2-ar-ratio}~(a) and (b).

We see that the inclusion of nuclear vibrations reduce the MO-SFA ratios by
a significant factor compared with the fixed nuclei calculations. The origin
of this effect was discussed in Sec.~\ref{sec:vibration}. The point is
simply that
the inclusion of vibrational motion will reduce the rate when the molecule
and the molecular ion have different bonding properties and transitions to
many different vibrational states occur.

\subsection{O$_2$:Xe}\label{sec:o2xe}

\begin{figure}
  \begin{center}
    \includegraphics[width=0.9\columnwidth]{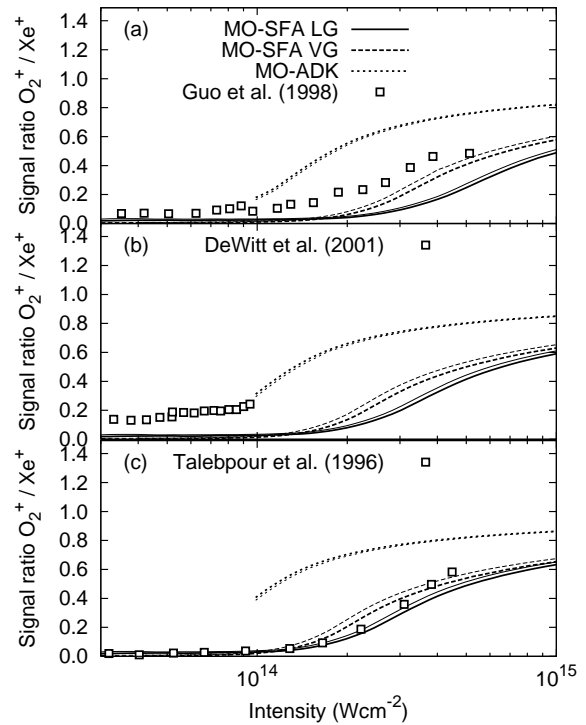}
  \end{center}
  \caption{Intensity dependent ratios between the yields of O$_2$ and Xe
  ions. In all panels the laser wavelength is $800\, \mbox{nm}$. 
  The thin lines are calculations with fixed nuclei and the thick 
  lines are the corresponding calculations including vibrations.
  In panel (a), the pulse
  duration is $30\, \mbox{fs}$ and the experimental data is from Ref.~\cite{Guo98}.
  In panel (b), the pulse duration is $100\, \mbox{fs}$ and
  the experimental data is from Ref.~\cite{DeWitt01}.
  In panel (c) the pulse duration is $220\, \mbox{fs}$ and
  the experimental data is from Ref.~\cite{talebpour96}.
}
  \label{fig:o2xe-ratios}
\end{figure}
Suppressed ionization of \mbox{O$_2$:Xe} has
been observed repeatedly~\cite{Guo98,DeWitt01,talebpour96}. 
Theoretically, this was explained in the MO-SFA  VG by the antibonding
character of the $\pi_g$ HOMO of O$_2$~\cite{muthbohm00}.
Another explanation within the MO-ADK
model was given in Ref.~\cite{Tong02} where the interpretation was based on
the asymptotic charge density of the $\pi_g$ HOMO. At some molecular
orientations the electronic density will be preferentially perpendicular to
the polarization axis and the rate of ionization will then be very small.
Finally in Ref.~\cite{Saenz00} it was proposed that nuclear dynamics could
be responsible for the suppression.
In Fig.~\ref{fig:o2xe-ratios}, we show the experimental data together with
our calculations. Clearly, all experiments and calculations show suppressed
ionization of O$_2$. Before saturation effects become important the
theoretical MO-SFA  VG predicts strongest suppression with the ratio below
$0.01$ while the ratio in MO-SFA  LG is $\approx 0.03$.
At low intensities, the experimental ratios are
scattered between $0.02-0.20$ which again makes a quantitative comparison
with theories difficult.
In the tunneling regime, beyond the intensity of $10^{14}\,
\mbox{Wcm}^{-2}$, the MO-ADK model predicts that saturation effects are already
important and thereby the degree of suppression is masked. 
Experimentally, saturation sets in at much higher intensities
around $2 \times 10^{14}\, \mbox{Wcm}^{-2}$, in good agreement with the MO-SFA
theories. 

\subsection{CO:Kr}\label{sec:cokr}

\begin{figure}
  \begin{center}
    \includegraphics[width=0.9\columnwidth]{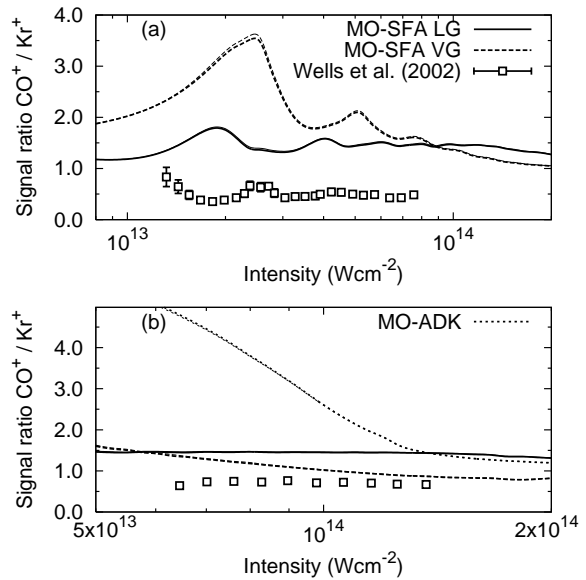}
  \end{center}
  \caption{Intensity dependent ratios between the yields of CO and Kr
  ions. In panel (a) the laser wavelength is $800\, \mbox{nm}$ and the pulse
  duration is $100\, \mbox{fs}$. In panel (b) the laser wavelength is 
  $1365\, \mbox{nm}$ and the pulse duration is $80\, \mbox{fs}$. In both
  panels the experimental data is from Ref.~\cite{Wells02}.
  The thin lines are calculations with fixed nuclei and the thick 
  lines are the corresponding calculations including vibrations.
  }
  \label{fig:cokr-ratios}
\end{figure}
The ion signal ratio of CO:Kr was measured in Ref.~\cite{Wells02} and
found to be around one half. In Figs.~\ref{fig:cokr-ratios}~(a) and
(b) we show the results at $800\, \mbox{nm}$ and $1365\,
\mbox{nm}$, respectively.
Under the experminental conditions at $800\, \mbox{nm}$, the Keldysh parameter 
exceeds unity. Thus we cannot rely on the MO-ADK theory and
in Fig.~\ref{fig:cokr-ratios}~(a) we show only MO-SFA calculations.
Both theories predict the ratio to be slightly larger than one. In the
MO-SFA VG model, we find that the ratio depends on the intensity in a
similar way as the N$_2$:Ar ratio, and since the electronic structures of CO and Kr are nearly identical
to N$_2$ and Ar, respectively,
we can explain the oscillating behaviour by the
same channel-closing argument (see Sec.~\ref{sec:n2ar}).

At $1365\, \mbox{nm}$ and in the intensity range of
Fig.~\ref{fig:cokr-ratios}~(b), we would expect to be in the tunneling
regime. Despite the fact that the tunneling theory should be applicable,
the MO-ADK model predicts the ratio to be somewhat too large. 
Finally we note that the result of
our MO-ADK calculation is approximately an order of magnitude larger than
the result from Ref.~\cite{Tong02}. The only differences between these two
calculations are the coefficients $C_{l, m=0}$ and the
time-averaging of the static field rate, our Eq.~(\ref{eqn:tunnelrate2}) as 
compared with Eq.~(10) of Ref.~\cite{Tong02}.

\subsection{Molecules without companion atoms}\label{sec:nocompanion}

\begin{figure}
  \begin{center}
    \includegraphics[width=0.9\columnwidth]{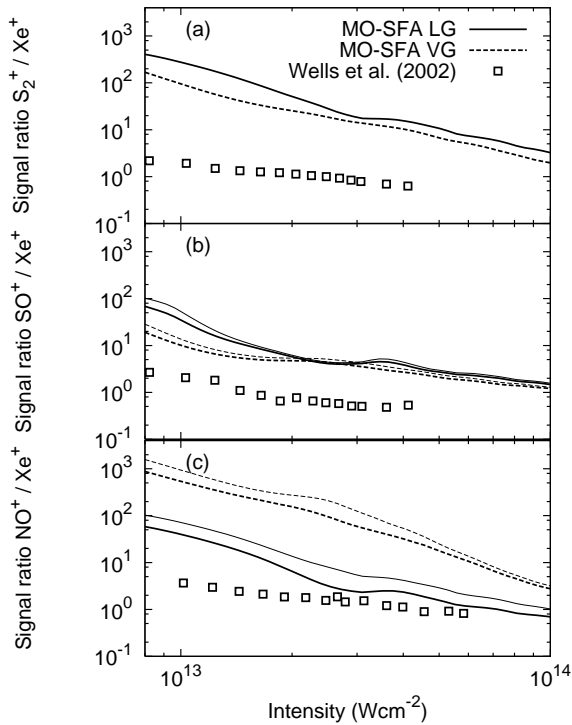}
  \end{center}
  \caption{
  Intensity dependent ratios between the ion yields of (a) S$_2$:Xe, (b)
  SO:Xe and (c) NO:Xe.
  In all panels the laser wavelength is $800\, \mbox{nm}$, the pulse
  duration is $100\, \mbox{fs}$ and the experimental data is from
  Ref.~\cite{Wells02}.
  The thin lines are calculations with fixed nuclei and the thick 
  lines are the corresponding calculations including vibrations.
}
  \label{fig:nocompaion-ratios}
\end{figure}

\begin{figure}
  \begin{center}
    \includegraphics[width=0.9\columnwidth]{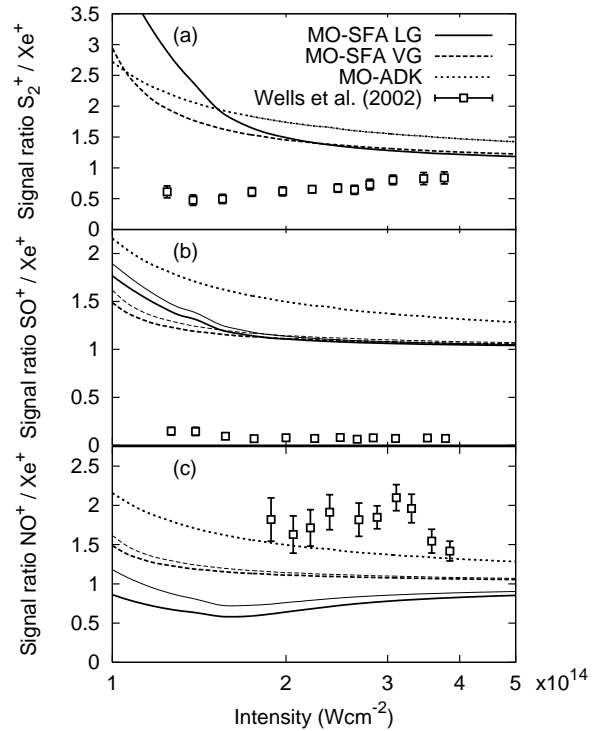}
  \end{center}
  \caption{
  Intensity dependent ratios between the ion yields of (a) S$_2$:Xe, (b)
  SO:Xe and (c) NO:Xe.
  In all panels the laser wavelength is $1365\, \mbox{nm}$, the pulse
  duration is $80\, \mbox{fs}$ and the experimental data is from
  Ref.~\cite{Wells02}.
  The thin lines are calculations with fixed nuclei and the thick 
  lines are the corresponding calculations including vibrations.
}
  \label{fig:nocompaion-1365}
\end{figure}

In Ref.~\cite{Wells02} the ion signal ratios of the pairs S$_2$:Xe, NO:Xe,
and SO:Xe were measured. Common to these pairs is that the ionization
potential of Xe is somewhat higher than the ionization potentials of the
molecules. The ratios measured are thus the results of both the structural
differences (orientation, electronic wave functions) 
and the difference between the binding energies.
The experiments were performed at the wavelength $800\, \mbox{nm}$ with
$\gamma > 1$, Fig.~\ref{fig:nocompaion-ratios}, and at $1365\, \mbox{nm}$
with $\gamma < 1$, Fig.~\ref{fig:nocompaion-1365}.

The electronic structure of S$_2$ is similar to O$_2$ and we should
therefore expect the ionization of S$_2$ to be suppressed too.
Our MO-SFA calculations, shown in Fig.~\ref{fig:nocompaion-ratios}~(a), predict the
S$_2$:Xe ratio to be much higher than measured.
Experimentally, the ratio was determined to be around unity. This indicates
that ionization of S$_2$ would indeed be suppressed if compared with a
hypothetical companion ion. 
In Figs.~\ref{fig:nocompaion-ratios}~(b)-(c) we show the calculations for
SO:Xe and NO:Xe and find a similar disagreement between theory and
experiment for these pairs.
The two previous attempts~\cite{Wells02,Tong02} to explain the ratios in this intensity regime 
overestimated the ratios by $3 - 5$ orders of magnitude.
Both calculations were based on tunneling models -- in Ref.~\cite{Wells02} 
a purely atomic ADK model was used and in Ref.~\cite{Tong02} the MO-ADK 
model was applied. Our present MO-SFA calculations including nuclear motion are significantly closer
to the experimental data but the agreement is still poor.

In Fig.~\ref{fig:nocompaion-1365} we show calculations and experimental data
at the wavelength of $1365\, \mbox{nm}$, i.e., in the tunneling regime. The
experiments were performed at such a high intensity that saturation effects
become important in all our calculations and correspondingly, the ratios are
all around unity. These predictions are in agreement with the experiment for
S$_2$:Xe, Fig.~\ref{fig:nocompaion-1365}~(a), and NO:Xe,
Fig.~\ref{fig:nocompaion-1365}~(c). From Fig.~\ref{fig:nocompaion-1365}~(b)
we see that the SO:Xe ratio is much lower than unity. This
is quite remarkable since all three ratios in
Fig.~\ref{fig:nocompaion-ratios} were nearly identical at $800\, \mbox{nm}$
but very different at $1365\, \mbox{nm}$  -- such
wavelength dependencies are of course impossible to predict by quasistatic 
tunneling theories. 

\section{Conclusions}
\label{Conclusions}

We have made a detailed study of strong-field ionization of diatomic
molecules and their companion atoms. In particular we have investigated
whether the approximate models: MO-SFA LG, MO-SFA VG and MO-ADK are
able to correctly predict the presence or absence of suppressed ionization 
in certain molecules.
Furthermore we have considered how nuclear vibrations can be
taken into account within the adiabatic theory and the 
single-active electron approximation.

All models did correctly predict absence of suppressed ionization for
N$_2$:Ar and presence of suppressed ionization for O$_2$:Xe.
Quantitative comparisons were made difficult due to disagreements between
different experiments, and further experiments would therefore be highly
desirable.

We find a rather good overall agreement between the MO-SFA LG theory 
and experiments on diatomic molecules and their companion atoms, and
we believe that the MO-SFA LG model accounts quite well for the 
structural dependency on the strong-field ionization rates. 
In general we found the length gauge MO-SFA to be in better agreement with
experiments than the velocity gauge MO-SFA -- in particular in the case of 
D$_2$:Ar where the MO-SFA VG predicted no suppression of ionization in
contradication with the experimental findings and the other theories.

The MO-ADK predictions were mostly in accordance
with previously published results using this model \cite{Tong02}.
For the cases N$_2$:Ar, D$_2$:Ar and O$_2$:Xe these predictions are also in
qualitative agreement with experimental data.

The most significant differences between the MO-ADK and the MO-SFA LG
calculations are the much lower ratios predicted by the latter theory in the
cases O$_2$:Xe and CO:Kr. In both cases the results of the MO-SFA LG model
agreed better with the experiments.
This is interesting in view of recent experiments
\cite{smits04} which reported suppression of strong-field ionization for
transition metal atoms relative to expectations, but only compared with
tunneling theory, and not SFA. For the future work it would be interesting to
check the SFA LG model on the transition metals to investigate if the
suppression is due to a general breakdown of the single-active electron
models or if it is due only to a failure of the ADK model.

All the models considered here fail to predict
the correct ratio between molecules and atoms with different ionization
potentials. This indicates that the dependency on the electronic
binding energy is not correctly accounted for and points to the need for the development of
improved models, including electron-electron correlations and exact final
states.

\begin{acknowledgments}
We would like to thank A. Saenz for useful discussions.
L.B.M. is supported by the Danish Natural Science Research Council (Grant
No. 21-03-0163).
\end{acknowledgments}

%\bibliography{sfabibfile}

\begin{thebibliography}{10}

\bibitem{muthbohm00}
J.~Muth-B\"{o}hm, A.~Becker, and F.~H.~M. Faisal,
\newblock Phys. Rev. Lett. {\bf 85}, 2280 (2000).

\bibitem{Kjeldsen04a}
T.~K. Kjeldsen and L.~B. Madsen,
\newblock J. Phys. B: At. Mol. Opt. Phys. {\bf 37}, 2033 (2004).

\bibitem{Tong02}
X.~M. Tong, Z.~X. Zhao, and C.~D. Lin,
\newblock Phys. Rev. A {\bf 66}, 033402 (2002).

\bibitem{ADK}
M.~V. Ammosov, N.~B. Delone, and V.~P. Krainov,
\newblock Sov. Phys. JETP {\bf 64}, 1191 (1986).

\bibitem{Keldysh}
L.~V. Keldysh,
\newblock Sov. Phys. JETP {\bf 20}, 1307 (1965).

\bibitem{Faisal}
F.~H.~M. Faisal,
\newblock J. Phys. B: At. Mol. Phys. {\bf 6}, L89 (1973).

\bibitem{Reiss80}
H.~R. Reiss,
\newblock Phys. Rev. A {\bf 22}, 1786 (1980).

\bibitem{talebpour96}
A.~Talebpour, C.~Y. Chien, and S.~L. Chin,
\newblock J. Phys. B: At. Mol. Opt. Phys. {\bf 29}, L677 (1996).

\bibitem{liang97}
Y.~Liang, A.~Talebpour, C.~Y. Chien, and S.~L. Chin,
\newblock J. Phys. B: At. Mol. Opt. Phys. {\bf 30}, 1369 (1997).

\bibitem{talebpour98}
A.~Talebpour, S.~Larochelle, and S.~L. Chin,
\newblock J. Phys. B: At. Mol. Opt. Phys. {\bf 31}, 2769 (1998).

\bibitem{Guo98}
C.~Guo, M.~Li, J.~P. Nibarger, and G.~N. Gibson,
\newblock Phys. Rev. A {\bf 58}, R4271 (1998).

\bibitem{DeWitt01}
M.~J. DeWitt, E.~Wells, and R.~R. Jones,
\newblock Phys. Rev. Lett. {\bf 87}, 153001 (2001).

\bibitem{Wells02}
E.~Wells, M.~J. DeWitt, and R.~R. Jones,
\newblock Phys. Rev. A {\bf 66}, 013409 (2002).

\bibitem{Helgaker00}
T.~Helgaker, P.~J\o{}rgensen, and J.~Olsen,
\newblock {\em Molecular Electronic-Structure Theory} (John Wiley \& Sons Ltd,
  Baffins Lane, Chichester, 2000).

\bibitem{Kjeldsen04c}
T.~K. Kjeldsen, C.~Z. Bisgaard, L.~B. Madsen, and H.~Stapelfeldt,
\newblock Phys. Rev. A (at print), e-print physics/0409150.

\bibitem{Chu01}
X.~Chu and S.-I.~Chu,
\newblock Phys. Rev. A {\bf 64}, 063404 (2001).

\bibitem{Otobe04}
T.~Otobe, K.~Yabana, and J.~I. Iwata,
\newblock Phys. Rev. A {\bf 69}, 053404 (2004).

\bibitem{Grasbon01}
F.~Grasbon {\em et~al.},
\newblock Phys. Rev. A {\bf 63}, 041402(R) (2001).

\bibitem{Gribakin97}
G.~F. Gribakin and M.~Y. Kuchiev,
\newblock Phys. Rev. A {\bf 55}, 3760 (1997).

\bibitem{condon30}
E.~U. Condon,
\newblock Phys. Rev. {\bf 36}, 1121 (1930).

\bibitem{Perelomov66}
A.~M. Perelomov, V.~S. Popov, and M.~V. Terent'ev,
\newblock Sov. Phys. JETP {\bf 23}, 924 (1966).

\bibitem{Becker01}
A.~Becker, A.~D. Bandrauk, and S.~L. Chin,
\newblock Chem. Phys. Lett. {\bf 343}, 345 (2001).

\bibitem{Smirnov66}
B.~M. Smirnov and M.~I. Chibisov,
\newblock Sov. Phys. JETP {\bf 22}, 585 (1966).

\bibitem{Bisgaard04}
C.~Z. Bisgaard and L.~B. Madsen,
\newblock Am. J. Phys. {\bf 72}, 249 (2004).

\bibitem{zare}
R.~N. Zare,
\newblock {\em Angular Momentum} (Wiley, New York, 1988).

\bibitem{webbook}
P.~J. Linstrom and W.~G. Mallard, editors,
\newblock {\em NIST Chemistry WebBook, NIST Standard Reference Database Number
  69} (National Institute of Standards and Technology, Gaithersburg MD, 20899,
  2003).

\bibitem{urbain04}
X.~Urbain {\em et~al.},
\newblock Phys. Rev. Lett. {\bf 92}, 163004 (2004).

\bibitem{kobus96}
J.~Kobus, L.~Laaksonen, and D.~Sundholm,
\newblock Comp. Phys. {\bf 98}, 346 (1996).

\bibitem{hf86}
C.~F. Fischer,
\newblock Comp. Phys. {\bf 43}, 355 (1986).

\bibitem{hf96}
G.~Gaigalas and C.~F. Fischer,
\newblock Comp. Phys. {\bf 98}, 255 (1996).

\bibitem{gribakin00}
G.~F. Gribakin, V.~K. Ivanov, A.~V. Korol, and M.~Y. Kuchiev,
\newblock J. Phys. B: At. Mol. Opt. Phys. {\bf 33}, 821 (2000).

\bibitem{dunn66}
G.~H. Dunn,
\newblock J. Chem. Phys. {\bf 44}, 2592 (1966).

\bibitem{halmann65}
M.~Halmann and I.~Laulicht,
\newblock J. Chem. Phys. {\bf 43}, 1503 (1965).

\bibitem{nicholls68}
R.~Nicholls,
\newblock J. Phys. B: At. Mol. Opt. Phys. {\bf 1}, 1192 (1968).

\bibitem{Berkowitz}
J.~Berkowitz,
\newblock J. Chem. Phys. {\bf 62}, 4074 (1975).

\bibitem{dyke97}
J.~Dyke {\em et~al.},
\newblock J. Chem. Phys. {\bf 106}, 821 (1997).

\bibitem{fittinghoff93}
B.~Chang, P.~R. Bolton, and D.~N. Fittinghoff,
\newblock Phys. Rev. A {\bf 47}, 4193 (1993).

\bibitem{jaron04}
A.~Jaro\'{n}-Becker, A.~Becker, and F.~H.~M. Faisal,
\newblock Phys. Rev. A {\bf 69}, 023410 (2004).

\bibitem{Saenz00}
A.~Saenz,
\newblock J. Phys. B: At. Mol. Opt. Phys. {\bf 33}, 4365 (2000).

\bibitem{smits04}
M.~Smits, C.~A. de~Lange, A.~Stolow, and D.~M. Rayner,
\newblock Phys. Rev. Lett. {\bf 93}, 213003 (2004).

\end{thebibliography}

\end{document}